# Concurrent growth and formation of electrically contacted monolayer transition metal dichalcogenides on bulk metallic patterns

*Sudiksha Khadka, Miles Lindquist, Shrouq Aleithan, Ari Blumer, Thushan Wickramasinghe, Martin Kordesch and Eric Stinaff ***

Sudiksha Khadka, Miles Lindquist, Shrouq Aleithan, Ari Blumer, Thushan Wickramasinghe, Martin Kordesch and Eric Stinaff
Department of Physics and Astronomy, Nanoscale and Quantum Phenomena Institute, Ohio University, Athens, Ohio, OH 45701, USA
E-mail: stinaff@ohio.edu

While new species and properties of two-dimensional (2D) materials are being reported with extraordinary regularity, a significant bottleneck in the field is the ability to controllably process material into working devices. We report a chemical vapor deposition based procedure to selectively grow 2D material in a deterministic manner around lithographically defined metallic patterns which concurrently provide as-grown contacts to the material. Monolayer films, with lateral extent of up to hundreds of microns are controllably grown on and around patterned regions of bulk transition metals. By using different combinations of metallic pattern and oxide precursor, heterostructured $MoS_2/WS_2$ growth has been observed as well. The materials display strong luminescence, monolayer Raman signatures, and relatively large crystal domains. In addition to producing high optical quality monolayer material deterministically and selectively over large regions, the metallic patterns have the advantage of providing as-grown contacts to the material, offering a path for device fabrication and large scale production.

The contemporary prominence of layered materials is driven by their technological and scientific potential in the 2D, monolayer, limit.[1,2] In addition to properties such as high mobilities, semiconducting and superconducting behavior, and excellent thermal properties, many of these materials have the potential for transformative opto-electronic applications, with large absorption, strong room-temperature emission, non-linear response, and optical control of spin and valley degrees of freedom.[3] Many seminal results of 2D material based device fabrication have involved the isolation of monolayers followed by making metal contact with the film, commonly using e-beam lithography. The customary method for isolating monolayers is micromechanical exfoliation which produces high quality crystalline flakes, on the order of up to tens of microns. However, this method provides no deterministic control over sample thickness, size, or location and provides no path to scalability. Liquid exfoliation uses ionic species as intercalating agents, which facilitates a breakdown of van der Waal forces and results in sub-micrometer sized monolayers of transition metal dichalcogenides (TMDs).[4,5] In addition to small sizes, liquid exfoliated monolayers are often found to have different structural and electronic properties, requiring further processing.[4] Chemical Vapor Deposition (CVD) has emerged as one of the most promising and preferred synthesis processes for TMD growth.[6] Two general methods of CVD growth include heating of a metal oxide powder in the presence of sulfur, [7] or direct sulfurization of thin layers of either metal or metal oxide.[8,9] Regardless of the technique used to produce monolayer TMDs, once they are grown or isolated, devices are then typically fabricated using lithography and metal deposition, which may require further processing in order to obtain useful metal-semiconductor contacts. Here we report a new technique, reversing this procedure, and growing TMDs directly on lithographically defined bulk metallic patterns. We have successfully produced uniform monolayer TMD growth in a deterministic area, on and between metallic patterns, with as-grown electrical contact to the material and high optical quality.

The 2D materials we focus on in this report are the well-known TMDs, MoS$_2$ and WS$_2$, which have diverse, and potentially useful, electronic, spintronic and optical properties.[10-19] The basic technique uses a standard lift-off lithography process along with DC sputtering to produce patterns composed of a given transition metal which then act as the nucleation and/or seed for the subsequent CVD growth. The CVD procedure involved sulfurization of the as-deposited metallic patterns in the presence of a small amount of a powder oxide precursor, either MoO$_3$ or WO$_2$, in addition to sulfur powder.[8,20,21] The overall procedure is relatively straightforward and robust providing a potential route for scalable fabrication of monolayer TMD based devices. The inset in **Figure 1a** shows an example four-probe structure of molybdenum on a Si/SiO$_2$ substrate where the Mo thickness was approximately 100 nm before growth. The metal patterns were sulfurized in the presence of MoO$_3$ powder at 780˚C for 15 mins in a continuous flow of Argon gas. As is seen in Figure 1a there is a continuous and uniform film growth around the metallic pattern. The width of the film is approximately 20 μm around the entire 58 mm perimeter of the device. A representative area of the film, highlighted by the red square in Fig. 1a, is analyzed in detail in Figure 1b through 1e. Atomic force microscopy, Figure 1c and 1f, shows an average monolayer thickness of 0.7 nm (Profile 1) as well as a distinct two layer step (Profile 2). This is consistent with the contrast on the optical image, Fig. 1b, and the photoluminescence (PL) spectra, Figure 1d, where the most intense peaks are centered around 660 nm and show a six-fold increase in intensity from two to one layer, comparable to typical exfoliated material.[14,15] As a final confirmation, Figure 1e, a plot of the difference in wavenumbers between the $E_{2g}^1$ and $A_{1g}$ lines in the Raman spectra, shows an absolute separation of 20 to 21 cm$^{-1}$ for the monolayer regions, consistent with previous CVD grown material.[22] Near the metal patterns the growth appears mostly multi-layer and as the film extends out the vast majority is comprised of monolayer material. A periodic array of islands is observed in the AFM scan at the edge of the film (Figure 1c and 1f). The exact nature of these islands is currently under investigation and may provide insight into the growth mechanism of the film. In this device pattern the probe separation is 4 μm and, within the region between the probes, the material was found to be predominantly multi-layer $MoS_2$. A subsequent series of growths were performed to optimize the production of monolayer material within the confined region between metallic features.

To produce monolayer material between metallic features with micron scale spacing it was necessary to reduce both the time and temperature of the growth process. **Figure 2** shows a representative region of $MoS_2$ grown on Mo patterns at 600˚C for 5 minutes where a continuous monolayer region can be found between the metallic patterns. Similar to Figure 1, the detailed optical analysis reported in Figure 2 shows characteristic monolayer photoluminescence and Raman signatures. The difference in intensity ratios of the A and B peak between Figure 1g and Figure 2d may indicate a physical, chemical, or structural difference between the monolayer regions in these two samples. A notable difference is that in Figure 1 the monolayer area is large and not bounded between two metal patterns which may lead to differences in strain profiles or defect density. The results in Figure 2a through 2f are representative of the larger area growth on the sample such as the 100 by 100 micron square scan, shown in Figure 2g through 2i, of the interdigitated molybdenum patterns where highly consistent and uniform monolayer growth is observed. Polarization resolved second harmonic generation imaging reveals that the material is polycrystalline with grain sizes on the order of a few microns.[23]

An added benefit of the reduced growth temperature is the reduced sulfurization of the sputtered metallic patterns. When samples were subjected to a 780˚C growth temperature it was found that the metal patterns would completely sulfurize and no longer be conductive, while at the lower temperature only a thin layer of the metal would sulfurize which was easily removed, revealing clean and highly conductive material underneath. Current vs voltage (IV) measurements show electrical contact with the as-grown monolayer material.[24,25] **Figure 3** shows an interdigitated electrode device structure which, after growth, results in an as-grown metal-semiconductor-metal Schottky barrier photodiode. The dark versus illuminated I-V curves show a clear photo-response qualitatively consistent with previous reports of $MoS_2$ based photodiodes.[26-32] The measured photocurrent ($I_{photocurrent} = I_{Illuminated} - I_{Dark}$), shown in Figure 3d, approaches 0.5 mA with a relatively symmetric response under uniform white-light illumination. These initial results on the first-generation as-grown devices demonstrate simplicity and potential of concurrent growth and formation of electrically contacted monolayer transition metal dichalcogenides on bulk metallic patterns.

Using one type of metal for the sputtered pattern and a different metal oxide powder resulted in the formation of heterostructured TMD material.[24] For example, in **Figure 4**, $MoS_2$ grown on tungsten patterns show Raman features as well as a red shift in the $MoS_2$ PL emission consistent with the formation of a $MoS_2$/$WS_2$ vertical heterostructure.[33,34] The Raman spectra in Figure 4 indicate few-layer growth for $MoS_2$ and multi-layer for $WS_2$ which would suppress other peaks observed by Gong *et al.*,[33] namely the low energy PL, attributed to a possible new direct exciton formation, as well as the $WS_2$ PL peak. For titanium patterns, as well as a platinum pattern done separately, we found no evidence of group VIB TMD formation,[24] indicating nucleation depends on not only topological features but chemical and materials properties of the deposited patterns as well.[35] These results also suggest that the metal patterns serve not only as a physical nucleation site for the initiation of the film, but act as a feedstock for initial and/or continued growth. To confirm this, a set of samples were grown without oxide precursors where it was found that direct sulfurization of the metallic pattern also resulted in monolayer TMD formation around the patterns.[24] A major difference between direct sulfurization and sulfurization in the presence of an oxide precursor is the lateral extent of the growth. Direct sulfurization appears to result in self-limited lateral films determined by the geometry and species of the initial metal pattern, whereas, with oxide precursors the growth may continue, fed by the precursor material. Therefore, deterministic patterns of complex multifunctional heterostructured materials may be grown through the use of mixed types of metallic patterns and transition metal/chalcogenide precursors.

The size, both laterally and vertically, of the initial metallic patterns is also found to affect the TMD formation. For growth on sub-micron metallic lines, as shown in Figure 4, TMD formation appears to be highly crystalline with large regions of monolayer material spanning tens to hundreds of microns. Again, the fact that the bulk metallic patterns remain after growth indicates that, though the pattern may be providing material to feed the growth, it is not entirely consumed, and the dominant supply of material is from the oxide precursor. Occasionally it was found that, after the photoresist removal and sample cleaning procedure, regions of the sputtered metal patterns would appear to be removed. However, patterned growth of high quality monolayer TMD material would often emerge in these regions,[24] indicating that very thin, optically transparent, layers of metal were still present in those regions. This is consistent with the findings of Gatensby, et al. and Woods, et al.,[8,34] where now, the film formation is predominantly constrained to the lithographically defined areas. This specificity coupled with the previous observation that small area growth tends to be crystalline could be used to produce wafer sized, lithographically defined, patterns of single-crystal regions suitable for subsequent device fabrication.

These results suggest that a general procedure of monolayer TMD growth on lithographically defined, bulk and/or thin film, metallic patterns can be tuned to produce good optical, as well as electronic, quality material over a range of dimensions from nanometers to hundreds of microns. In fact it is expected that as the feature sizes reduce the crystal domains will be on the order of the device, possibly forming the basis of large scale incorporation of monolayer based TMD materials in opto-electronic device applications. Instead of trying to start with a large area of high quality material and then creating devices by, for example, patterning, contacting, and etching, we envision a process whereby material is controllably and selectively grown only between patterned metallic regions. In addition to being relatively straightforward and scalable, this method provides as-grown contacts to the material with consistent quality, high crystallinity due to smaller growth area, the ability to easily mix materials, and straightforward incorporation into existing semiconductor device fabrication techniques.

**Experimental Section**

*Sample growth.* Metallic patterns on Si/SiO$_2$ substrates are produced via a standard lift-off procedure. Patterns are made using an image reversal photoresist (Clariant AZ-5214E) and a home built contact optical lithography rig which is followed by DC sputtering of the metal (Denton Vacuum DV-502A), and removal of the photoresist. The substrates, with patterns, are cleaned using DI water, Acetone and Isopropanol. A graphite boat is used to hold the oxide powder with the substrate placed upside down above the oxide region on the holder. For MoS$_2$ growths, Molybdenum Trioxide, ACS, 99.5+%, CAS: 1313-27-5, from Chemsavers was used. For WS$_2$ growths, Tungsten(IV) Oxide, 99.9%, CAS: 12036-22-5, from Chemsavers was used. The graphite boat with oxide powder and sample is then placed toward the closed end of a one inch diameter, 14 in long, half open, quartz tube. A boron nitride boat with sulfur powder (Sublimed Sulfur, CAS: 7704-34-9, from Spectrum Chemicals) is place inside the same tube at the open end. This tube is then placed inside a larger, 2 inch, tube inside a Barnstead Thermolyne Digital Lab Tube Furnace, Model F21135, such that the sample is in the middle of the furnace and the sulfur is outside where external heater tape is used to produce the sulfur vapor. The larger tube is closed and a flow of argon gas inside the main tube is maintained in the direction of the open to closed end of the smaller tube.[24] The ratio, and amounts, of oxide and sulfur powder are based on the feature size of metallic pattern or the intended area of growth of semiconductor film.

For growth at 780˚C, the furnace temperature is raised to 550˚C, at the rate of 35˚C/min; to 650˚C, at the rate of 15˚C/min and then to final growth temperature of 780˚C, at the rate of 4˚C/min. For growth at 600˚C, the furnace temperature is raised to 550˚C, at the rate of 35˚C/min; to 600˚C, at the rate of 15˚C/min. Once the growth temperature is reached sulfur is evaporated at about 200˚C using the external heater and is allowed to enter into the growth chamber. The start of the growth time is measured from the point at which sulfur starts to melt. At the end of the growth period, the furnace is turned off and left to cool down to room temperature.

*Sample characterization.* Well established methods for the characterization of as-grown materials were used, including, optical contrast imaging, photoluminescence (PL) spectroscopy, Raman spectroscopy, AFM imaging and polarization resolve second harmonic imaging. Optical imaging, with color contrast of the sample, was carried out using a Navitar MicroMate video microscope. PL and Raman spectra were collected using a WITec α-SNOM300s microscope where the excitation source was a 532 nm cw-dpss single-mode fiber coupled laser and the power was kept less than 1 mW. AFM scans were performed using a Thermomicroscopes Autoprobe CP-Research in non-contact mode. Polarization resolved second harmonic generation imaging was done to measure crystallinity and grain size using a pulsed Ti-Sapphire (SpectraPhysics Tsunami) producing 2 ps pulses (80MHz rep rate) at 830 nm and an energy per pulse of approximately 5 μJ (tens of mW average power).[23] I-V curves were recorded using a Keithley model 2400 sourcemeter. Contact to the device pads was made using a conductive epoxy.

**Supporting Information**
Supporting Information is available online from the Wiley Online Library or from the author.

**Acknowledgements**
This work was supported by the Ohio University CMSS and NQPI programs. The authors would like to thank Prof. Hugh Richardson for access to, and assistance with, the Witec system.

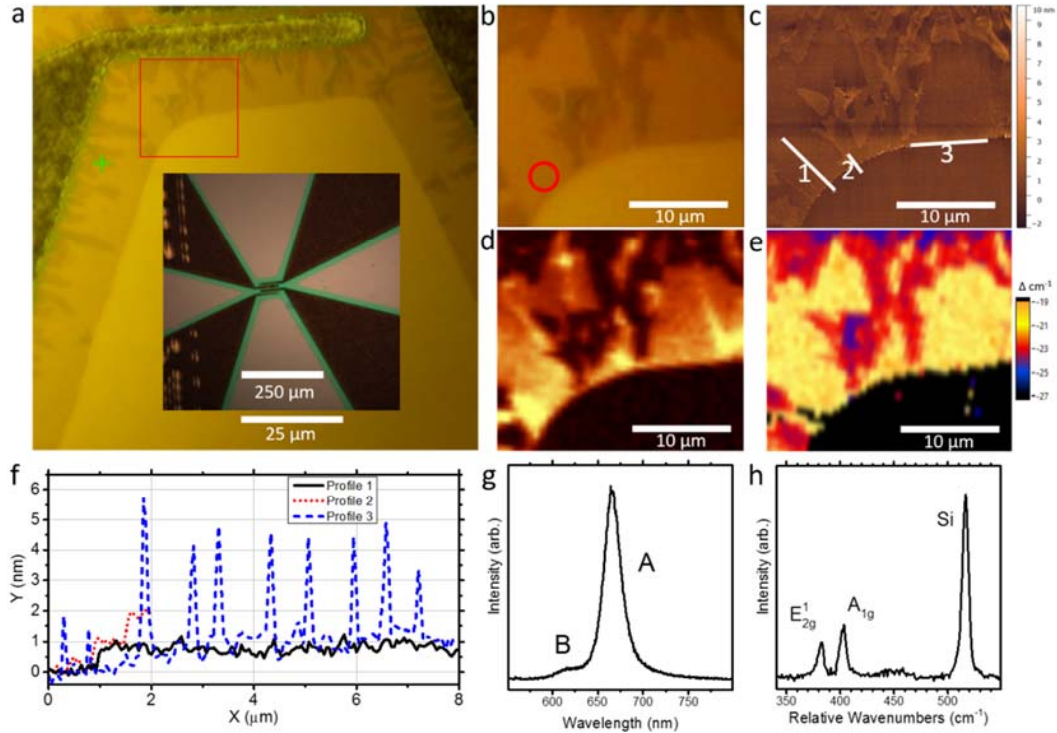

**Figure 1.** Optical and AFM analysis of MoS$_2$ grown on a molybdenum 4-probe device. **a,** Large area optical image of the as-grown sample where the inset shows a continuous uniform film around the edge of the Mo pattern. **b-e**, Higher magnification optical image, AFM scan, photoluminescence (PL) map, and Raman analysis map of the region highlighted by the red square in **a**. The brightness of the PL map is proportional to the integrated intensity between 550 to 750 nm, while the color scale on the Raman map indicates the separation between $E^1_{2g}$ and $A_{1g}$ in wavenumbers. **f**, Line profiles extracted from the AFM scan in **c** where Profile 1, black solid line, shows a uniform film with thickness less than 1 nm. Line Profile 2, red dotted line, shows a two layer region. Line Profile 3, blue dashed line, is taken along an edge of the film where a periodic array of islands is observed. The islands are ~150 nm in diameter and ~4 nm in height and occur regularly along the edge with a periodicity on the order of a micron. **g and h**, Photoluminescence and Raman spectra from the circled region in **d**.

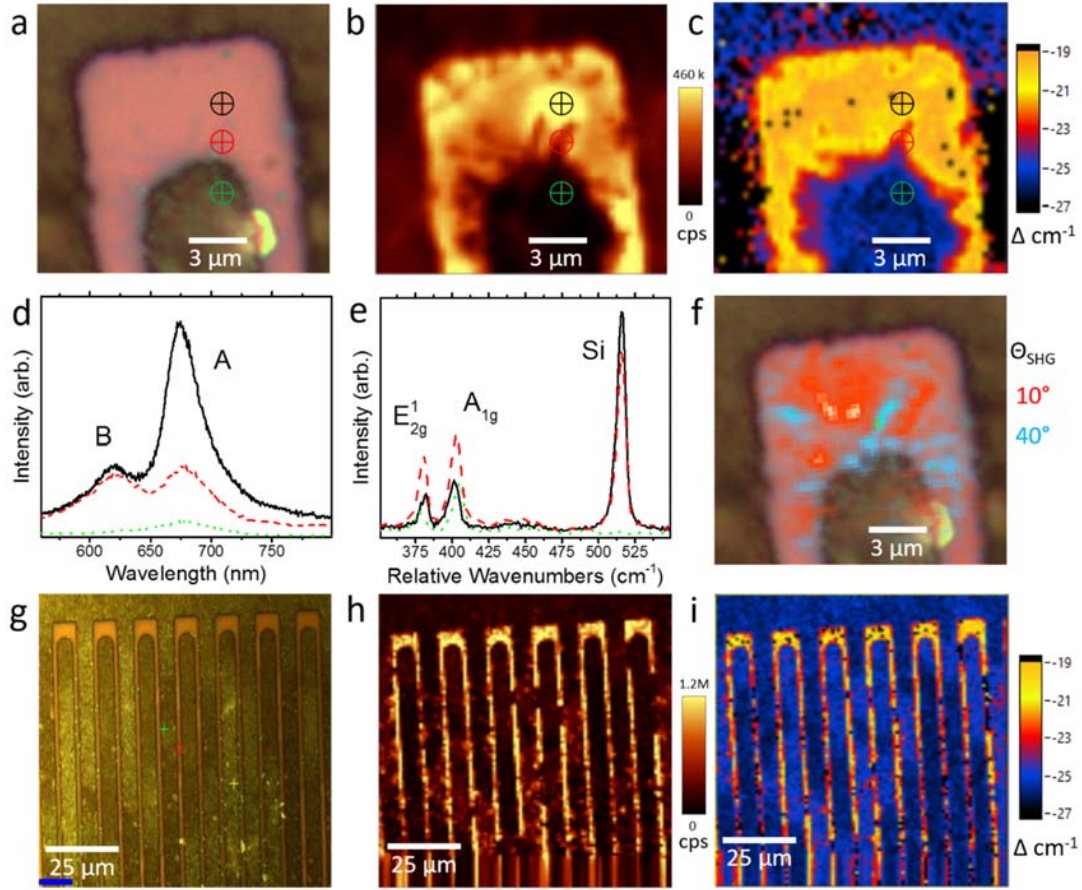

**Figure 2.** Optical analysis of continuous monolayer MoS$_2$ film formation between molybdenum features. **a-c,** Optical image, photoluminescence map, and Raman analysis map, respectively. Similar to Figure 1, the brightness of the PL map is proportional to the integrated intensity between 550 to 750 nm, while the color scale on the Raman map indicates the separation between the $E_{2g}^1$ and $A_{1g}$ in wavenumbers. **d and e**, Photoluminescence and Raman spectra from the regions indicated with the black, red and green targets. From these spectra it can be seen that the MoS$_2$ covers the metallic patterns as well as the region between. There is a continuous monolayer film in the region between the Mo patterns. **f,** Polarization resolved second harmonic generation (SHG) imaging indicating crystalline grains on the order of a few microns. **g-i,** Optical image, photoluminescence map, and Raman analysis map, similar to **a-c**, for a 100 by 100 micron square region.

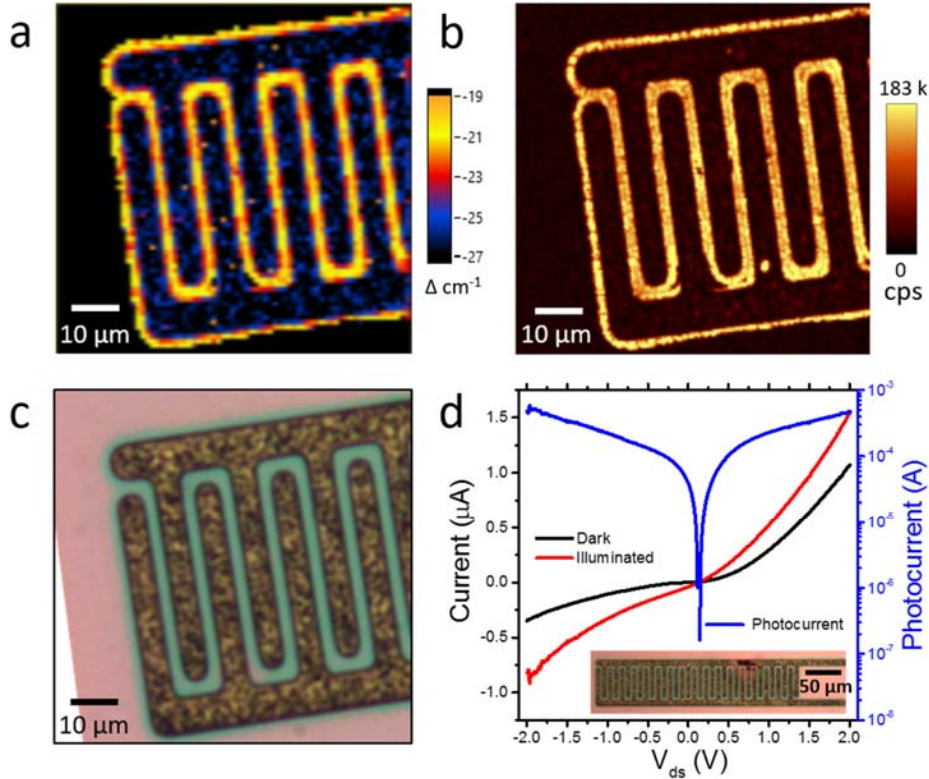

**Figure 3.** Opto-electronic analysis of as-grown monolayer MoS$_2$ metal-semiconductor-metal photodiode. **a-c,** Raman analysis map, photoluminescence map, and optical image respectively. Similar to Figure 1, the brightness of the PL map is proportional to the integrated intensity between 550 to 750 nm, while the color scale on the Raman map indicates the separation between the $E^1_{2g}$ and $A_{1g}$ in wavenumbers. **d,** Current versus voltage measurements of the device with and without illumination from an unfocused broadband white light led with a power density of approximately 100 pW/μm$^2$. The inset in **d** shows the active device region with uniform monolayer coverage.

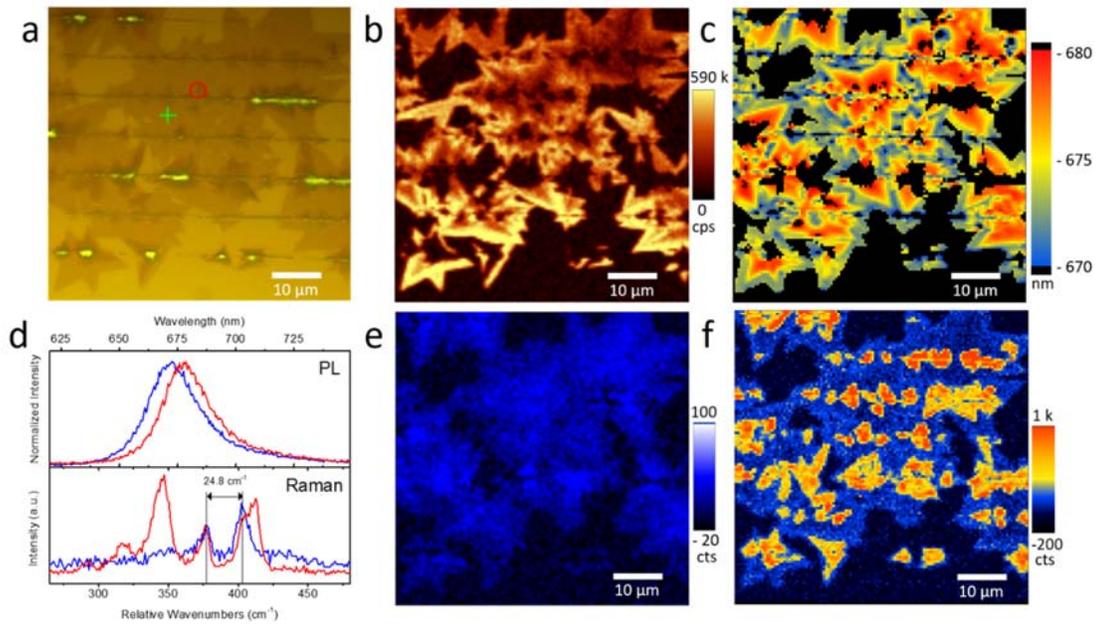

**Figure 4.** Heterostructured MoS₂/WS₂ material grown on thin tungsten metallic lines. **a,b,** Optical and total PL intensity images. **c,** Peak position of a Gaussian fit to the PL emission. A blue shift is observed toward the edges of the material. **d,** PL and Raman spectra corresponding to the red and blue targets on **a,b and c**. **e,** Integrated Raman intensity of the $E^1_{2g}$ and $A_{1g}$ lines of MoS₂. **f,** Integrated Raman intensity of the $E^1_{2g} + 2LA$ and $A_{1g}$ lines of WS₂. The observation of the WS₂ Raman lines in the middle of the material is consistent with red shifted PL peak in this region.

**Concurrent growth and formation of electrically contacted monolayer transition metal dichalcogenides on bulk metallic patterns**

*Sudiksha Khadka, Miles Lindquist, Shrouq Aleithan, Ari Blumer, Thushan Wickramasinghe, Martin Kordesch and Eric Stinaff \**

**S1. Direct sulfurization of metallic patterns.** Samples grown without oxide powder still displayed film growth extending from the metallic patterns. **Figure S1** shows an example of a molybdenum metallic pad subjected to only sulfur vapor at various growth temperatures. This indicates the metallic patterns serve as both a nucleation site as well as a feedstock for limited growth. Temperatures below 600°C display no extended film growth and the initial metallic patterns show evidence of sulfurization. For growths at 600°C, the metal shows initiation of growth at the edges leaving the bulk metal slightly sulfurized. At 780°C it is found that the metal pad is heavily sulfurized forming an extended film of $MoS_2$. This temperature also results in complete transformation of the metallic pattern into bulk $MoS_2$. At the highest temperature multi-layer formation is also observed in the extended film. The film appears to provide an active growth site for subsequent layers.

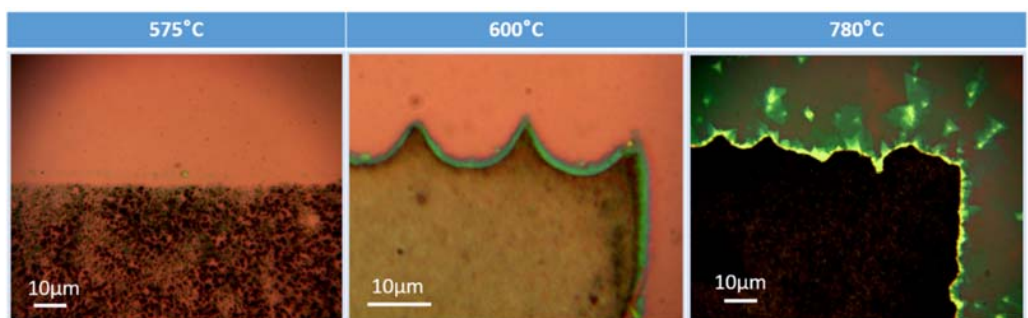

**Figure S1.** Direct sulfurization of molybdenum patterns. Temperature dependent sulfurization of a molybdenum metallic pattern. Film formation is observed to extend from the pattern for temperatures at or above 600°C.

**S2. Thin metal film growth.** Unintentional variations in our home-built photolithography system resulted in regions where the sputtered metal widths were sub-micron and occasionally it was found that, after the photoresist removal and sample cleaning procedure, regions of the sputtered metal patterns would appear to be removed. As seen in **Figure S2**, patterned growth of high quality monolayer TMD material would often emerge in these regions, indicating that very thin, optically transparent, layers of metal were still present. This is consistent with the findings of Gatensby, et al. and Woods, et al.,[1,2] where now the film formation is predominantly constrained to the lithographically defined areas.

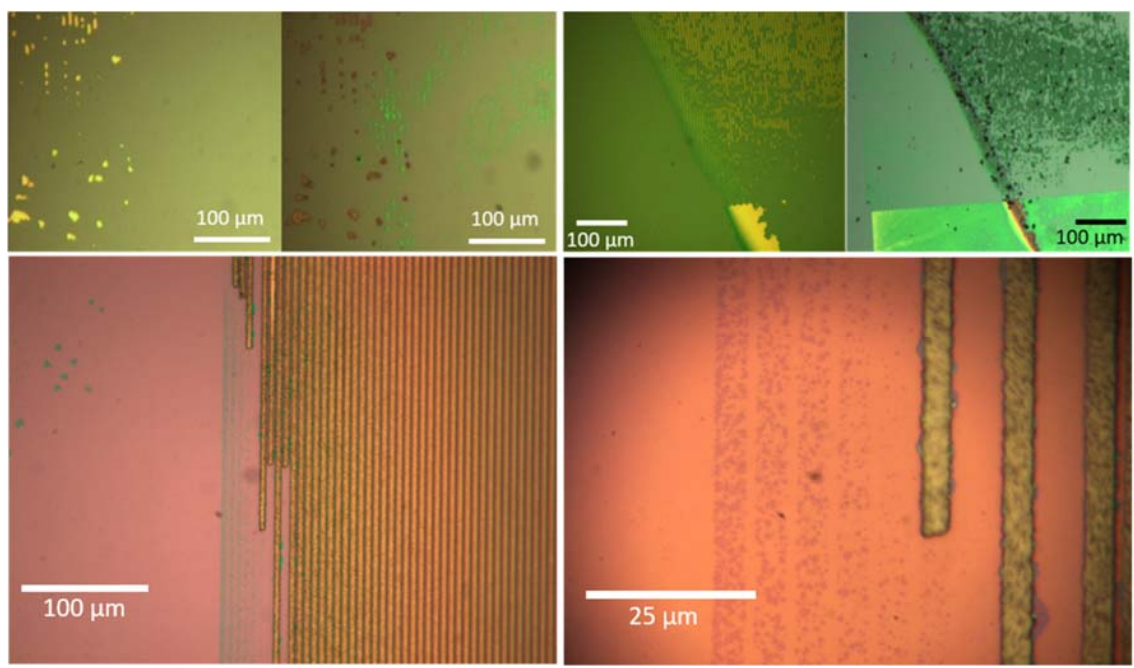

**Figure S2.** TMD film growth on thin metal films. Film formation on monolayer, optically transparent, sputtered metal is observed. Here the growth is confined to the lithographically defined regions.

**S3. MoS$_2$ growth on Ti. Figure S3** shows an optical image of a Ti pattern subjected to a MoS$_2$ growth procedure. No evidence of group VIB TMD formation was observed in PL and Raman experiments. The optical, PL and Raman scans on a Ti pattern subjected to a WS$_2$ growth showed similar results.

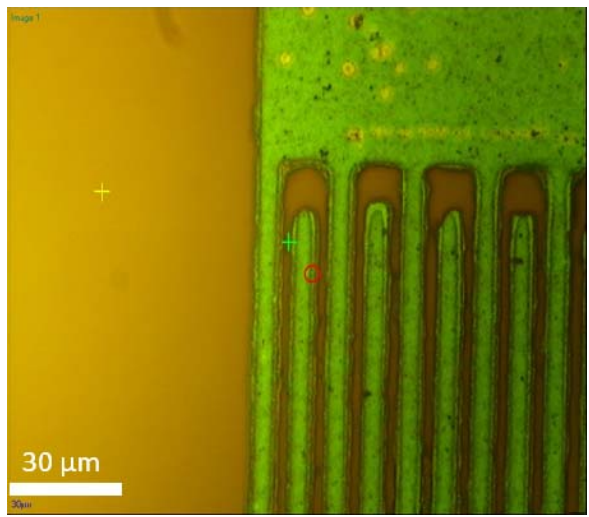

**Figure S3.** Ti pattern subjected to MoS$_2$ growth procedure.

**S4. Series of TMD film growths on various metallic features.** To investigate the role of the sputtered metal species on the growth of $MoS_2$ and $WS_2$, patterns of Molybdenum, Tungsten and Titanium were sputtered together on a pair of substrates. The substrates were then placed in the CVD furnace and subjected to identical procedures in the presence of $MoO_3$ and $WO_2$ powder. The results indicate that the type of metal deposited is a critical parameter in the formation of the TMD films. In the four combinations shown in **Figure S4**, luminescent TMD film growth is observed and while the inter-species growths ($MoS_2$ on W and $WS_2$ on Mo) appear the most intense they also show the broadest emission as well as showing Raman signatures of hybrid, $MoS_2/WS_2$, material formation. The Raman analysis maps, as well as the relatively weak PL intensities, indicate predominantly multi-layer material for these growth conditions.

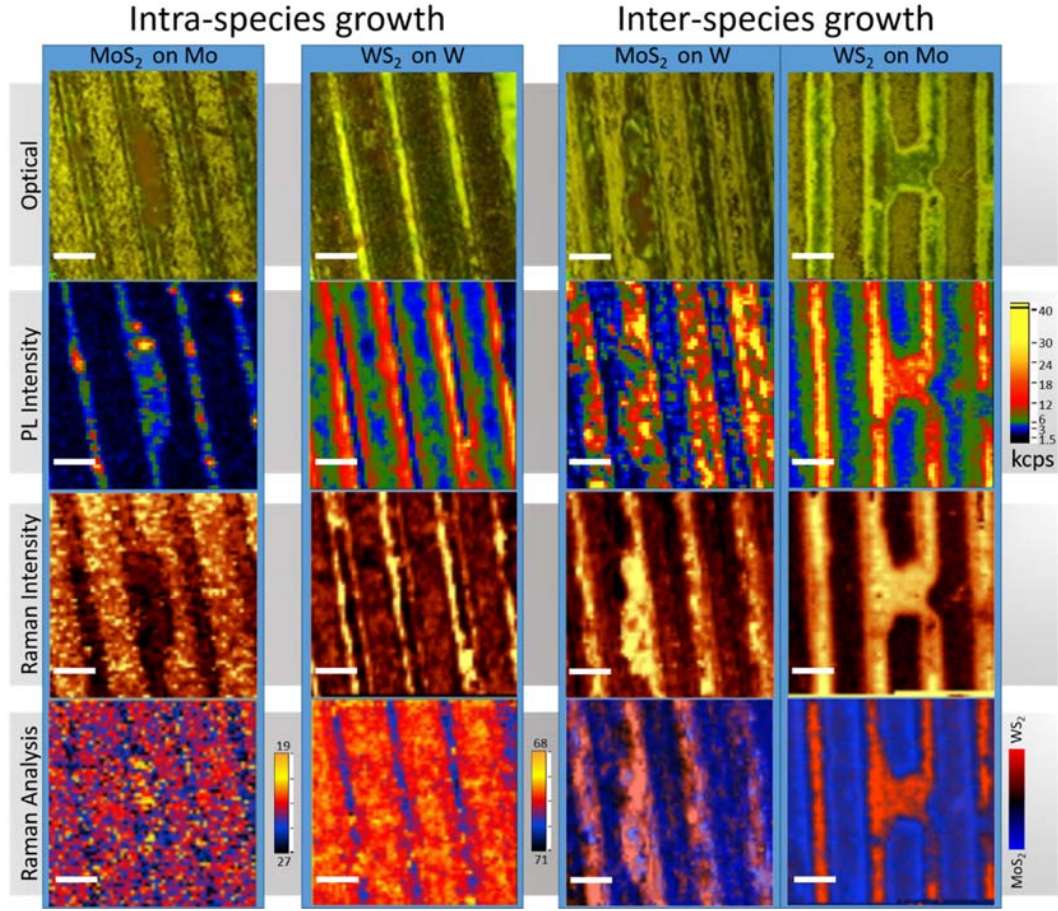

**Figure S4.** Optical, PL and Raman (intensity and analysis) images of TMD film formation on two types of sputtered metallic patterns (molybdenum and tungsten). The scale bar for all images is 6 μm. Growth on titanium patterns (Figure S3) resulted in no detectable PL or Raman signatures of group VIB TMD material. The PL intensity maps are proportional to the integrated intensity between 550 to 750 nm. The color scale applies to all PL maps and has a non-uniform color spacing to highlight the large intensity differences. The Raman analysis maps for the intra-species growths (MoS$_2$ on Mo and WS$_2$ on W) indicate the separation between the $E^1_{2g}$ and $A_{1g}$ lines for MoS$_2$, and the $E^1_{2g} + 2LA$ and $A_{1g}$ lines for WS$_2$, in wavenumbers. The Raman analysis maps for the inter-species growths (MoS$_2$ on W and WS$_2$ on Mo) indicate the integrated intensity of the $E^1_{2g}$ and $A_{1g}$ lines for MoS$_2$ (blue) and the $E_{2g}^1 + 2LA$ and $A_{1g}$ lines for WS$_2$ (red) and show the formation of heterostructured MoS$_2$/WS$_2$.

**S5. WS₂ growth on W.** An example growth of WS₂ is shown in **Figure S5**. This was grown on W patterns in the presence of WO₂ powder at a temperature of 780°C. The PL emission was taken under similar conditions to those in the main text and displays a higher total emission.

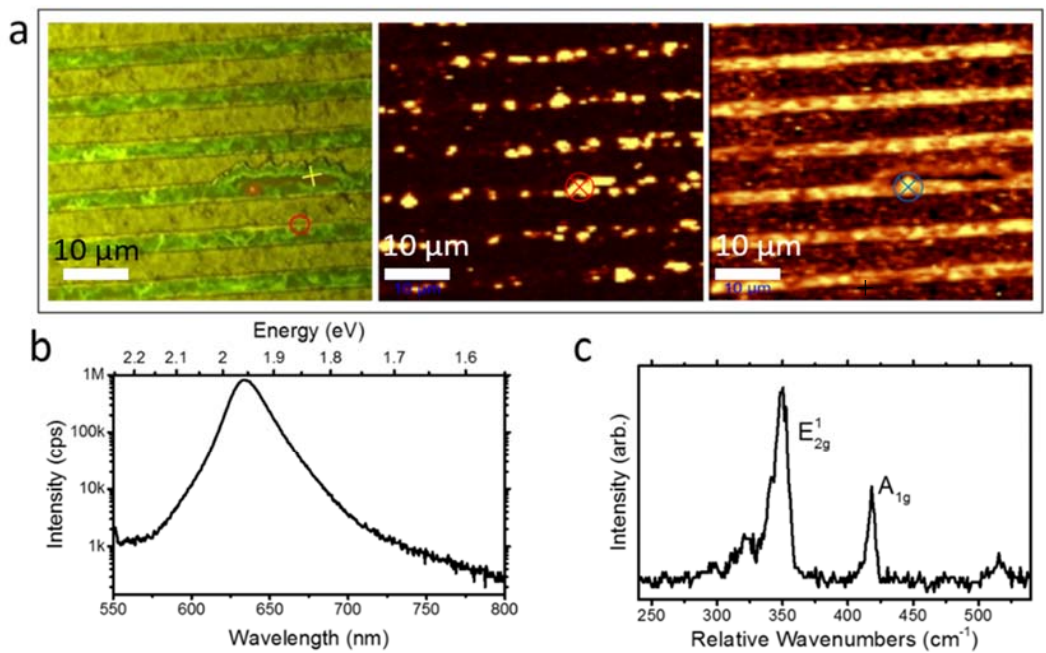

**Figure S5.** WS₂ grown on W patterns. a, Optical, PL and Raman intensity and analysis images of WS₂ grown in the presence of WO₂ powder at 780°C. The PL intensity map is proportional to the integrated intensity between 550 to 750 nm and the Raman intensity map is proportional to the integrated Raman intensity of the $E_{2g}^1 + 2LA$ and $A_{1g}$ lines of WS₂. b, PL spectrum from the region marked by the red target in the PL intensity map. c, Raman spectrum from the region marked by the blue target in the Raman intensity map.

**S6. Time dependent growth series. Figure S6** shows three growth durations for MoS$_2$ grown on molybdenum patterns at 600°C in the presence of MoO$_3$ powder.

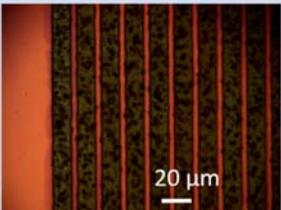

| Growth time | Optical Images | Observation on growth pattern |
|---|---|---|
| 1.5 min | 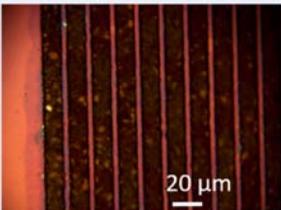 | 1. Uniform thin film of MoS$_2$.<br>2. Area of growth is small as compared to that of 5 min growth. |
| 5 min | | 1. Uniform thin film dominated by monolayer region.<br>2. Area of growth is uniform in wider area of the substrate. |
| 15 min | 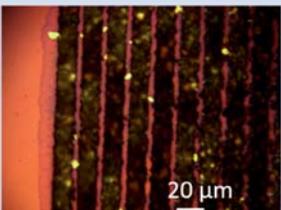 | 1. Not a uniform growth. Has few mono layered film area. Mostly they are thicker as compared to 5min of growth.<br>2. Area of growth is much larger in comparison to that of 5 min of growth. |

**Figure S6.** Time dependent growth series. Optical images for MoS$_2$ growth on Mo patterns for different growth durations measured from the initiation of the sulfur evaporation.

**S7. Schematic of the cvd set up.** The CVD set up used for the growths is shown in **Figure S7**.

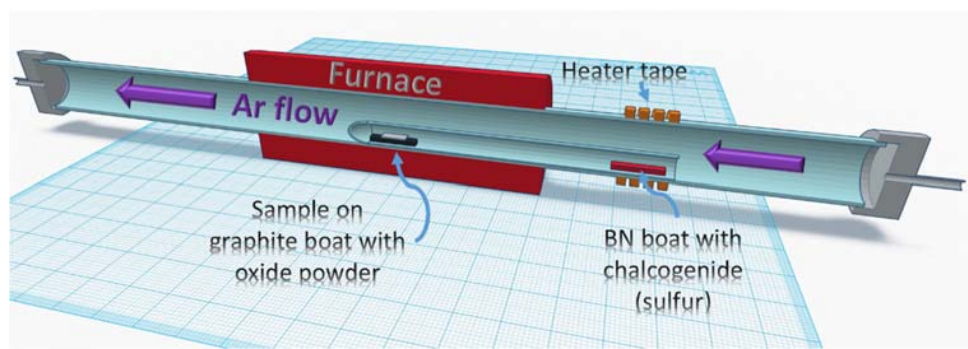

**Figure S7.** Schematic diagram of the CVD set up.